\documentclass[proceedings, preprint]{rmaa}

\usepackage{paralist}

\usepackage{psfrag,color}

\SetYear{2008}
\SetConfTitle{Magnetic Fields in the Universe II}

\title{Studying ISM magnetic fields and turbulent regimes from polarimetric maps} 

\author{
  D. Falceta-Gon\c calves,\altaffilmark{1,2} 
  A. Lazarian,\altaffilmark{2}
  and G. Kowal\altaffilmark{3}}

\altaffiltext{1}{N\' ucleo de Astrof\' isica Te\' orica, Universidade Cruzeiro do
Sul - Rua Galv\~ ao Bueno 868, CEP 01506-000, S\~ao Paulo, Brazil 
(diego.goncalves@unicsul.br).}

\altaffiltext{2}{Astronomy Department, University of Wisconsin,
 Madison, 475 N. Charter St., WI 53711, USA.}

\altaffiltext{3}{Astronomical Observatory, Jagiellonian University, ul. Orla 171, 30-244 Krak\'ow, Poland.}

\shortauthor{Falceta-Gon\c calves, Lazarian, \& Kowal}
\shorttitle{Turbulence from polarization maps of MHD simulations}

\listofauthors{D. Falceta-Gon\c calves, A. Lazarian, \& G. Kowal}
\indexauthor{Falceta-Gon\c calves, D.}
\indexauthor{Lazarian, A.}
\indexauthor{Kowal, G.}

\abstract{Polarimetric maps have been used for the characterization of 
the magnetic field in molecular clouds. However, it is difficult to determine the 
3-dimensional properties of these regions from the projected maps. For
that reason, numerical simulations can be used as benchmarks for polarimetric measurements, 
and eventually reveal more about the interplay of turbulence and the magnetic 
field lines. In this work we make a number of MHD numerical simulations of turbulent 
molecular clouds and created their synthetic dust emission polarization maps, 
varying the direction of the observer. We determined the correlation of 
intensity emitted and polarization degree for the simulated models. We were able to reproduce 
the decay of the polarization degree at denser regions without any assumption regarding the 
properties of the dusty component. The anti-correlation arises from the simple 
cancellation of the polarization vectors along the line of sight. This effect is 
amplified 
within denser regions as the magnetic field configuration becomes more complex. We 
studied the probability distribution function, the power spectrum, and the structure function of 
the polarization angles. This statistical analysis revealed strong differences depending 
on the turbulent regime (i.e. sub/supersonic and sub/super-Alfvenic). Therefore, these 
methods can be used on polarimetric observations to characterize the dynamics of 
molecular clouds. We also presented a modified Chandrashekhar-Fermi method to obtain the 
intensity of the local magnetic field. The proposed formulation showed no limitations 
regarding orientation or turbulent regime.}

\resumen{
Mapas polarim\'etricos han sido usados para caracterizar el campo
magn\'etico en nubes moleculares. Sin embargo, es dif\'{\i}cil
determinar las propiedades tridimensionales de mapas de proyecci\'on
de dichas regiones. Por esta raz\'on, simulaciones num\'ericas pueden
ser usadas como pruebas de las mediciones polarim\'etricas, y
eventualmente revelar m\'as acerca de la relaci\'on entre la
turbulencia y las l\'{\i}neas de campo magn\'etico. En este trabajo
realizamos una serie de simulaciones MHD de nubes moleculares
turbulentas y creamos mapas sint\'eticos de emisi\'on polarizada de
polvo, en las cuales variamos la direcci\'on del
observador. Determinamos la correlaci\'on de la intensidad de la
emisi\'on y el grado de polarizaci\'on para los modelos. Pudimos
reproducir el decaimiento del grado de polarizaci\'on en zonas m\'as
densas sin suposici\'on alguna de las propiedades de la componente de
polvo. La anticorrelaci\'on viene la simple cancelaci\'on de los
vectores de polarizaci\'on a lo largo de la visual. El efecto es
amplificado dentro de zonas de alta densidad ya que la configuraci\'on
del campo magn\'etico se torna m\'as compleja. Estudiamos la funci\'on
de distribuci\'on de probabilidad, el espectro de potencias, y
funciones de estructura de los \'angulos de polarizaci\'on. Estos
an\'alisis estad\'\i sticos revelan gandres diferencias dependiendo
del r\'egimen turbulento (es decir sub \'o s\'uper-s\'onico, sub \'o
super-Alfv\'enico). Entonces, estos m\'etodos pueden ser utilizados en
observaciones polarim\'etricas para caracterizar la din\'amica de las
nubes moleculares. Presentamos tambi\'en un m\'etodo
Chandrassekhar-Fermi modificado para obtener la magnitud del campo
magn\'etico local. La formulaci\'on sugerida no muestra limitaciones
respecto al r\'egimen turbulento u orientaci\'on.
}

\addkeyword{ISM: magnetic fields}
\addkeyword{techniques: polarimetric}
\addkeyword{methods: numerical}
\addkeyword{methods: statistical}

\begin{document}
% Typeset article header
\maketitle

\section{Introduction}
\label{sec:intro}

Giant molecular clouds in the interstellar medium (ISM) are believed to be threaded by 
large scale magnetic fields \citep{sch98, cru99}. 
However, it is still not completely clear what is the role of the magnetic field in the dynamics of the ISM 
and what is its effect on the star formation process. Also, the ratio of the magnetic and turbulent energy in these 
environments is a subject of controversy \citep{padoan02, girart06}.
Polarimetric maps have been extensively used for the determination of the 
magnetic fields in several astrophysical environments. This technique represents the 
best tool to characterize the vector components of the magnetic field 
parallel to plane of the sky. For galaxies and the intracluster medium this technique can  
be used for polarization of synchrotron emission \citep{hil00}.

For a given polarization map of an observed region, the mean polarization angle indicates the orientation of the 
large scale magnetic field. On the other hand the polarization dispersion 
gives clues on the value of the turbulent energy. This, as a consequence, can be used to determine the magnetic field component in the 
plane of sky. Chandrasekhar \& Fermi (1953) introduced a method (CF method hereafter) for 
estimating the ISM magnetic fields 
based on the dispersions of the polarization angle and gas velocity. Simply, it is assumed that the magnetic field 
perturbations are Alfvenic and that the rms velocity is isotropic.

A promising approach to test this method is to create two-dimensional (plane of sky) synthetic 
maps from numerically simulated cubes. 
Ostriker, Stone \& Gammie (2001) performed 3D-MHD simulations, with $256^3$ resolution, 
in order to obtain polarization maps and study the validity 
of the CF method on the estimation of the magnetic field component along the plane of sky. They showed that the CF method gives 
reasonable results for highly magnetized media, in which the dispersion of the polarization angle is $< 25^{\circ}$. However, 
they did not present any other statistical analysis or predictions that could be useful to determine the 
ISM magnetic field from observations. Polarization maps from numerical simulations can also be used in the study of the correlation between the polarization 
degree and the total emission intensity (or dust column density). Observationally, the polarization degree in dense 
molecular clouds decreases with the total intensity as $P \propto I^{-\alpha}$, with $\alpha = 0.5 - 1.2$ \citep{goncalves05}.
Padoan et al. (2001) studied the role of turbulent cells in the 
$P \ vs \ I$ relation using supersonic and super-Alfvenic
self-gravitating MHD simulations. They found a decrease of the polarization degree 
with total dust emission within gravitational cores, in agreement with observations, if grains are assumed to be unaligned for $A_V > 3$. When the 
alignment was assumed to be independent on $A_V$, the anti-correlation was not observed. 
Recently, Pelkonen, Juvela \& Padoan (2007) extended this work and 
refined the calculation of polarization degree introducing the radiative transfer properly. In that work, the decrease in 
the alignment efficiency arises without any {\it ad hoc} assumption. The alignment efficiency decreases as the radiative 
torques become less important in the denser regions. However, it is 
still not clear the role of the magnetic field topology and the presence of multiple cores intercepted by the line of sight 
on the decrease of polarization degree.

In this work we attempt to extend the previously cited studies improving and applying the CF method to different 
situations. For that, we used both sub and super-Alfvenic models, to study the role of the magnetic field topology in the 
observed polarization maps.

\section{Numerical Simulations}
\label{sec:numerics}

The simulations were performed solving the set of ideal MHD equations, in conservative form, as follows:

\begin{equation}
\frac{\partial \rho}{\partial t} + \mathbf{\nabla} \cdot (\rho{\bf v}) = 0,
\end{equation}

\begin{equation}
\frac{\partial \rho {\bf v}}{\partial t} + \mathbf{\nabla} \cdot \left[ \rho{\bf v v} + \left( p+\frac{B^2}{8 \pi} \right) {\bf I} - \frac{1}{4 \pi}{\bf B B} \right] = \rho{\bf f},
\end{equation}

\begin{equation}
\frac{\partial \mathbf{B}}{\partial t} - \mathbf{\nabla \times (v \times B)} = 0,
\end{equation}

\noindent
with $\mathbf{\nabla \cdot B} = 0$, where $\rho$, ${\bf v}$ and $p$ are the plasma density, velocity and pressure, 
respectively, ${\bf B}$ is the magnetic field and ${\bf f}$ represents the external acceleration source, responsible for 
the turbulence injection. For molecular clouds, we may assume that the ratio of dynamical to radiative timescales is very large. Under this assumption, 
the set of equations is closed with an isothermal equation of state $p = c_s^2 \rho$, where 
$c_s$ is the speed of sound. The equations are solved using a second-order-accurate and non-oscillatory 
scheme, with periodic boundaries, as described in Kowal, Lazarian \& Beresniak (2007).

Initially, we set the intensity of the x-directed magnetic field ${\bf B_{ext}}$ and the gas thermal pressure $p$. This allows us to obtain 
sub-Alfvenic or super-Alfvenic, and subsonic or supersonic models.

The turbulent energy is injected using a random solenoidal function for ${\bf f}$ in Fourier space. This, 
in order to minimize the influence of the forcing in the formation of density structures. We inject energy at scales $k \propto L/l<4$, where $L$ 
is the box size and $l$ is the eddy size of the injection scale. The rms velocity $\delta V$ is kept close to unity, therefore ${\bf v}$ and the Alfv\'en speed $v_A = 
B/\sqrt{4 \pi \rho}$ will be measured in terms of the rms $\delta V$. Also, the time $t$ is measured in terms of the dynamical 
timescale of the largest turbulent eddy ($\sim L/\delta V$). 

We performed four computationally extensive 3D MHD simulations, using high resolution ($512^3$), for different initial conditions, as shown in Table 1. 
We simulated the clouds up to $t_{\rm max} \sim 5$, i.e.\ 5 times longer than the dynamical timescale, 
to ensure a full development of the turbulent cascade. 
We obtained one subsonic and three supersonic models. One of the supersonic models is also super-Alfvenic. Each data cube contains information 
about parameterized density, velocity and magnetic field. As noted from Eqs. (1) and (2), the simulations 
are non self-gravitating and, for this reason, the results are scale-independent.

Regarding the gas distribution in each model we found an increasing
contrast as we go to higher sonic Mach number, independently on 
the Alfvenic Mach number. Subsonic turbulence 
show a gaussian distribution of densities, while the increased number and strength of 
shocks in supersonic cases create 
smaller and denser structures. In these cases, the density contrast may be increased by a factor of 100 - 10000 compared 
to the subsonic case. The magnetic field topology, on the other hand, depends on the Alfvenic Mach number. Sub-Alfvenic models 
show a strong uniformity of the field lines, while the super-Alfvenic case shows a very complex structure. Both effects, the 
density contrast and the magnetic field topology, may play a role on the polarimetric maps, as shown further in the paper.

\begin{table}
\begin{center}
\caption{Description of the simulations\label{table1}}
\begin{tabular}{cccccc}
\hline\hline
Model & $P$ & $B_{ext}$ & Description \\
1 &1.00 &1.00 & subsonic \& sub-Alfvenic\\
2 &0.10 &1.00 & supersonic \& sub-Alfvenic\\
3 &0.01 &1.00 & supersonic \& sub-Alfvenic\\
4 &0.01 &0.10 & supersonic \& super-Alfvenic\\
\hline\hline
\end{tabular}
\end{center}
\end{table}

\section{Results}
\label{sec:results}

\subsection{Polarization Maps}
\label{sec:maps}

To create the polarization maps we assumed the dust polarization to be completely 
efficient and that the radiation is originated exclusively by thermal emission from 
perfectly aligned grains. Under 
these assumptions, the local angle of alignment ($\psi$) is determined by the local magnetic field projected into the 
plane of sky, and the linear polarization Stokes parameters $Q$ and $U$ are given by: 

\begin{eqnarray}
q = \rho \cos 2\psi \sin^2 i, \nonumber \\
u = \rho \sin 2\psi \sin^2 i,
\end{eqnarray}

\begin{figure*}
   \centering
   \includegraphics[width=5cm]{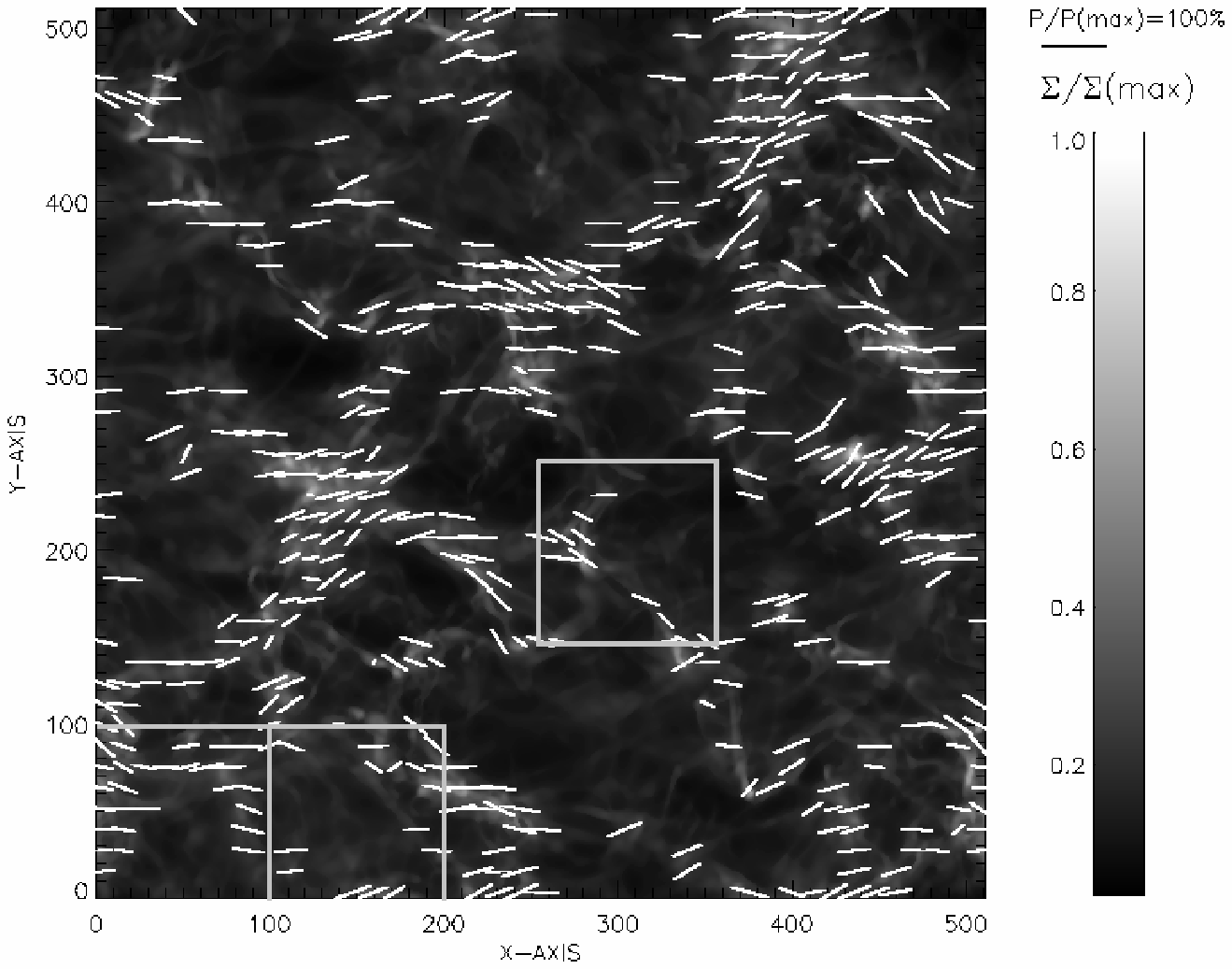}
   \hspace{0.2cm}
   \includegraphics[width=5cm]{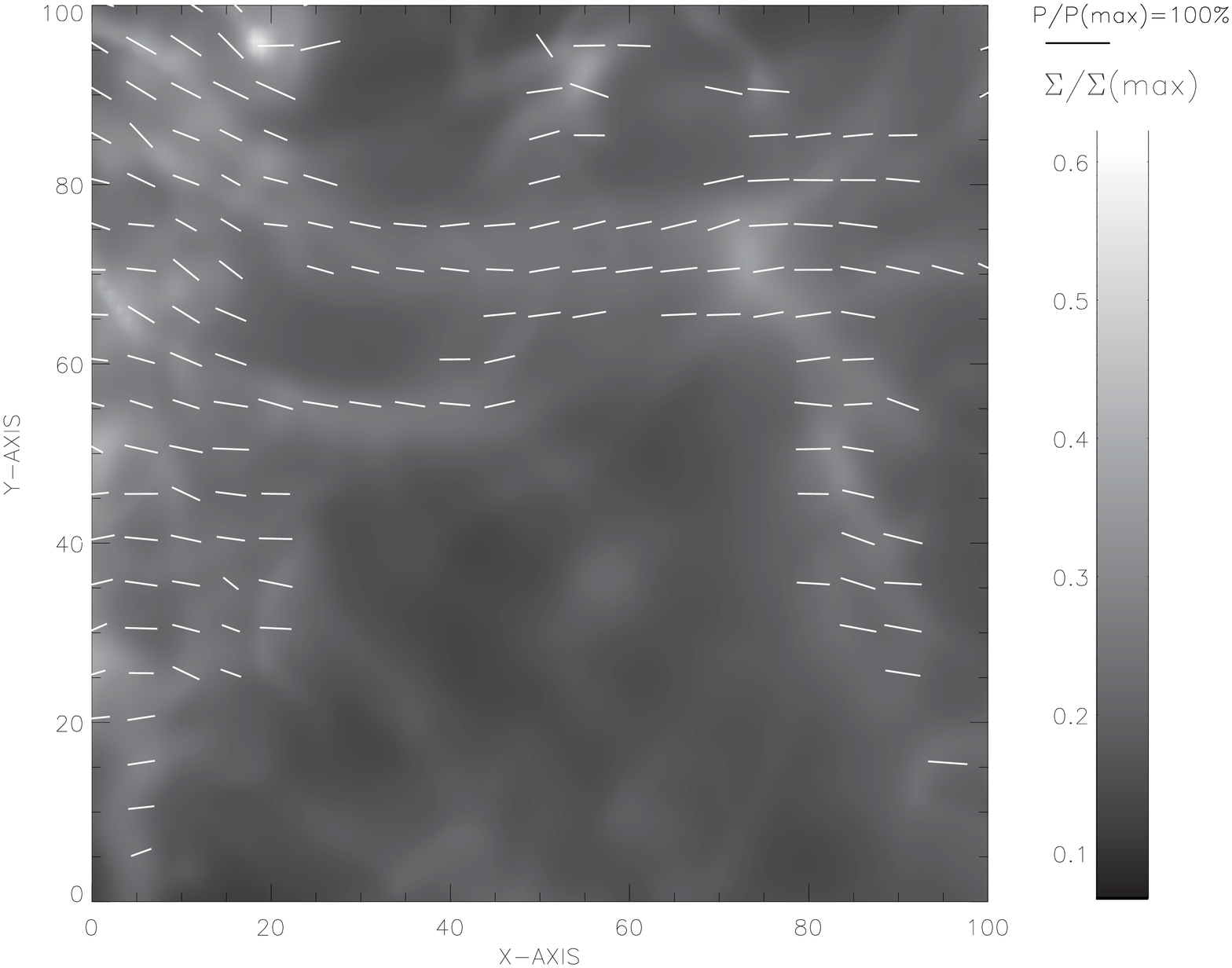}\\[20pt]
   \includegraphics[width=5cm]{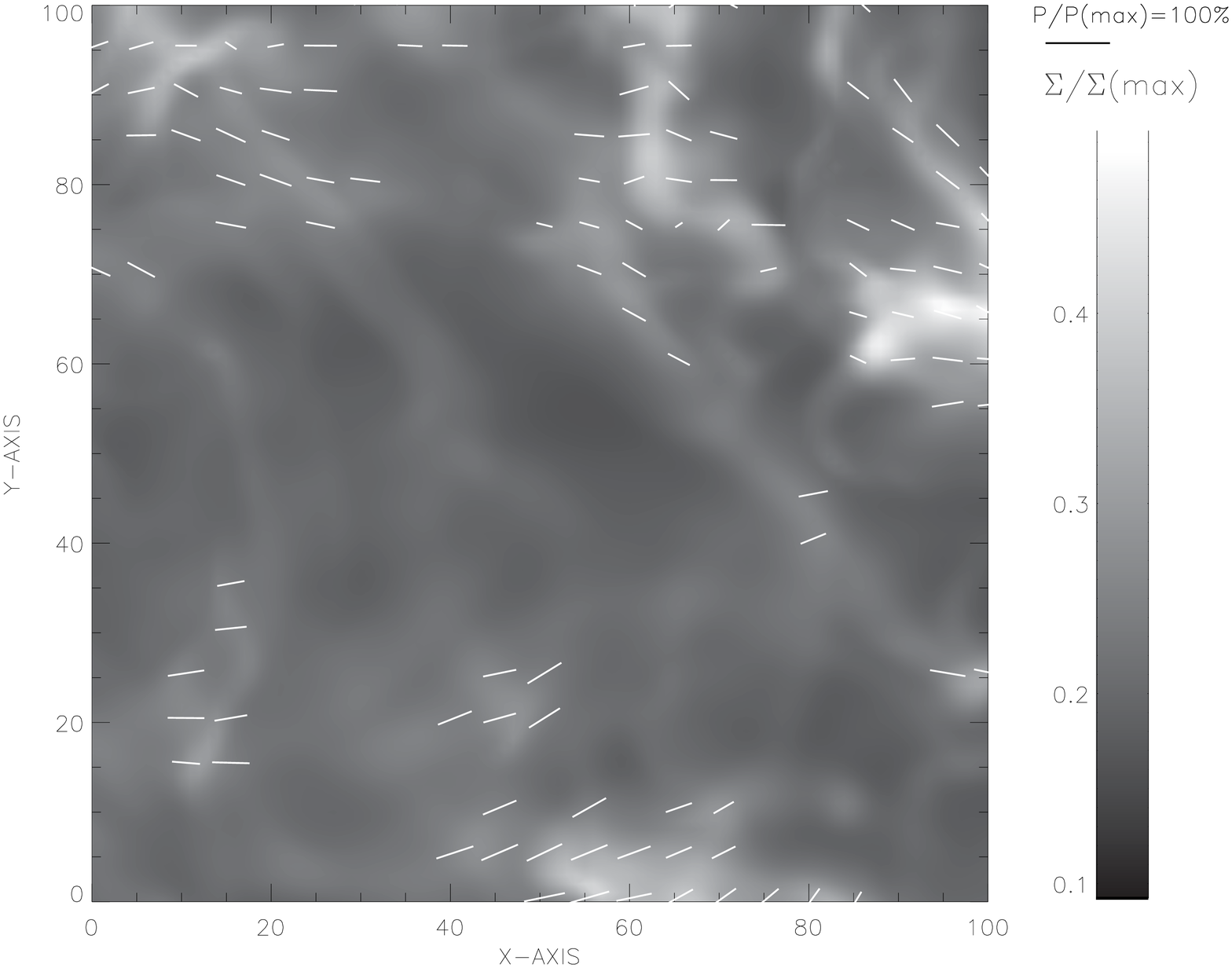}
   \hspace{0.2cm}
   \includegraphics[width=5cm]{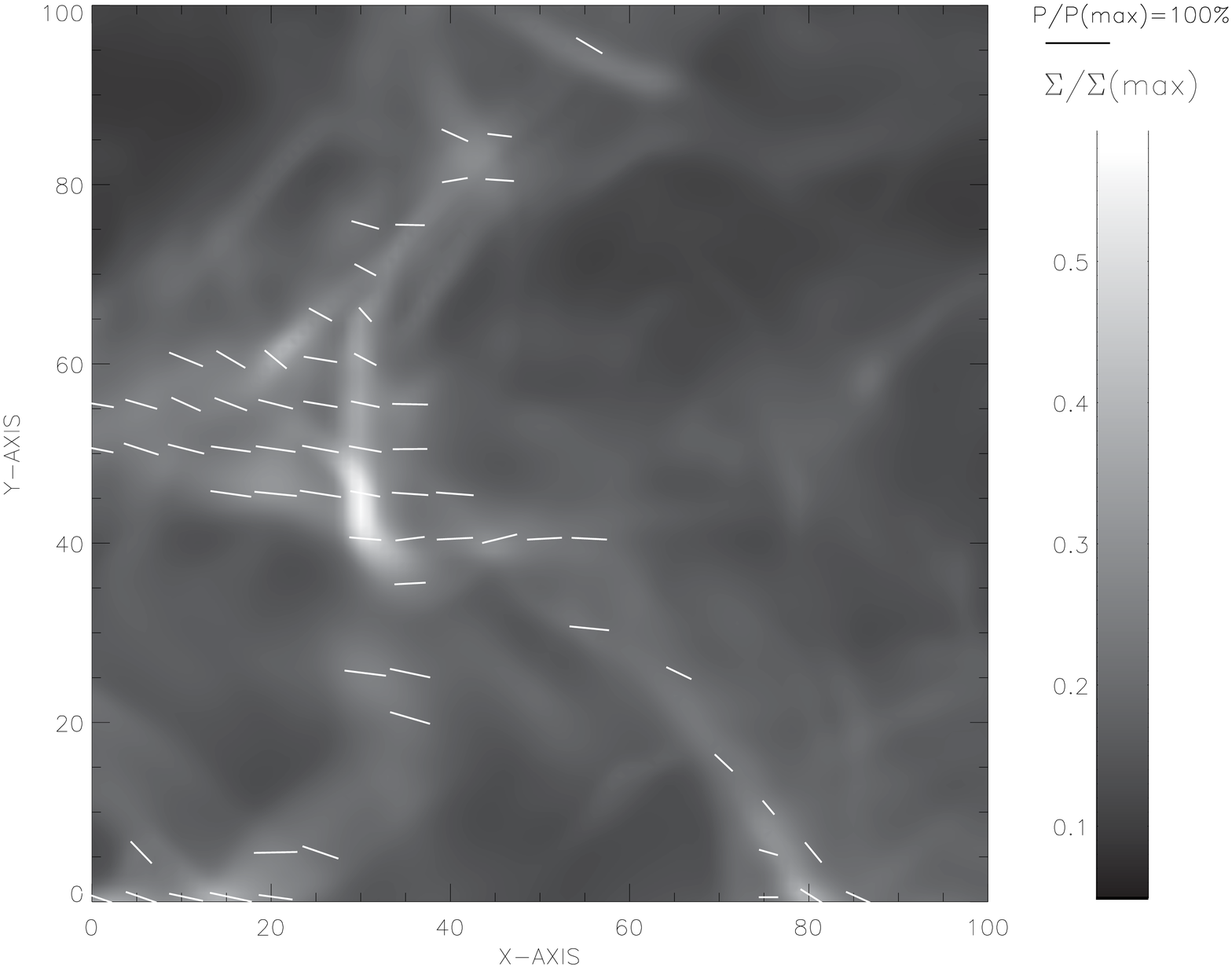}
   \caption{Polarization of emission and column density maps for Model 3 ($M_{\rm S} \sim 7.0$  and $M_{\rm A} \sim 0.7$) with ${\bf B}_{ext}$ perpendicular to the line of sight. The complete map (512x512 pixels) ({\it Upper-left}) and the zoomed regions (100x100 pixels). The sensitivity in simulated observations is assumed to be 0.3 of the maximum emission. Here, regions where the signal is less than 0.3 do not show polarization vectors, and $P_{\rm max} = 97\%$.}
\end{figure*} 

\begin{figure*}
   \centering
   \includegraphics[width=5cm]{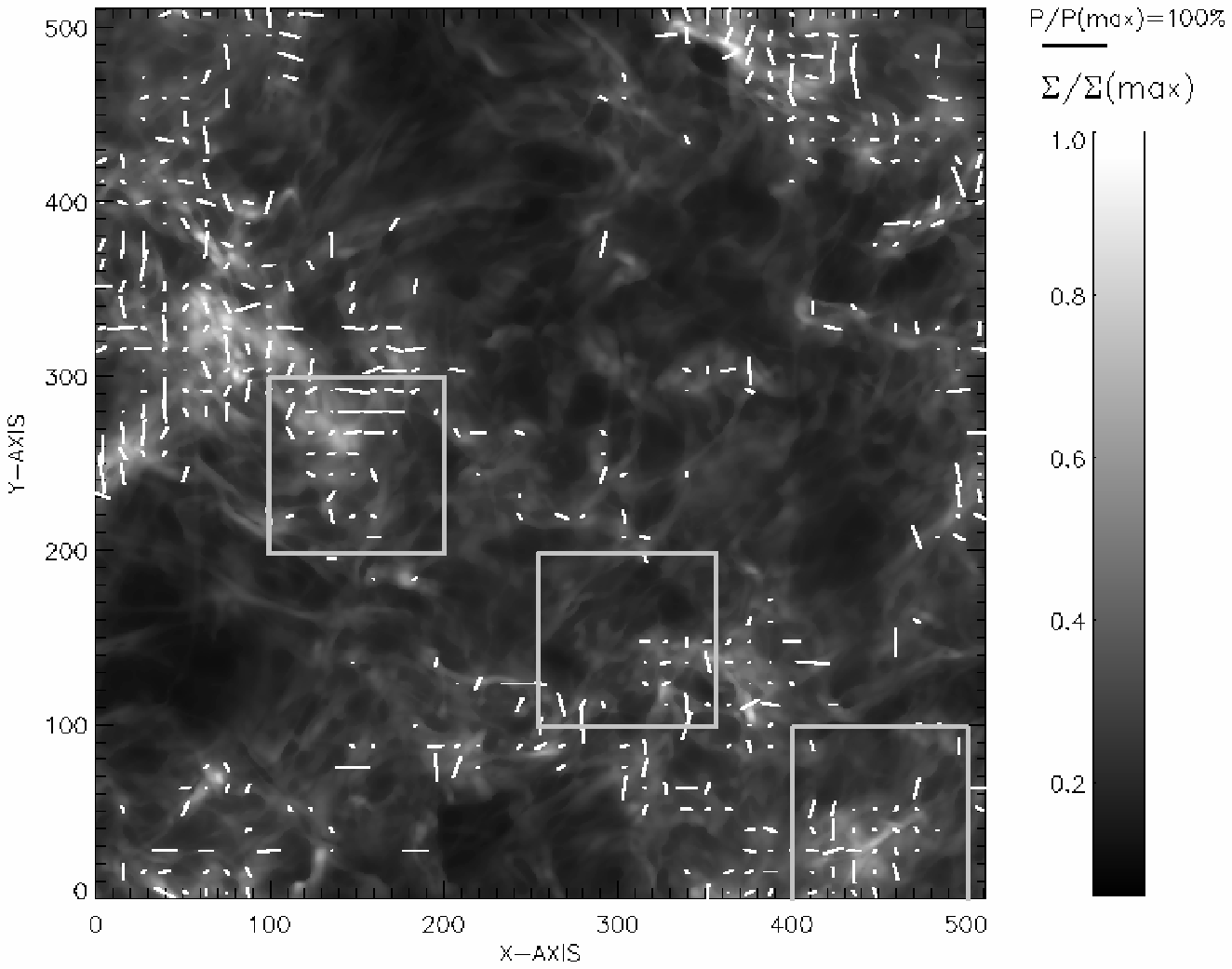}
   \hspace{0.2cm}
   \includegraphics[width=5cm]{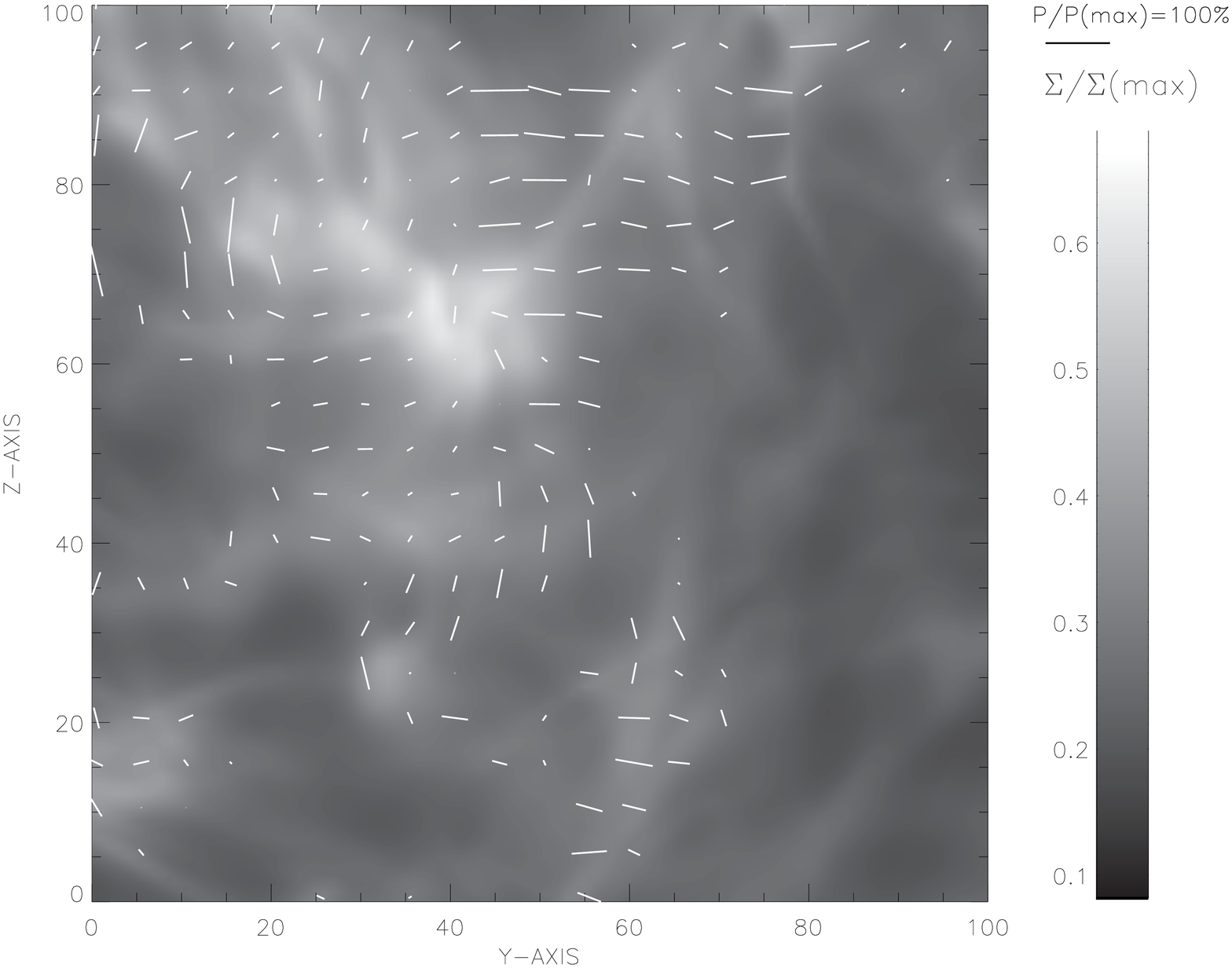}\\[20pt]
   \includegraphics[width=5cm]{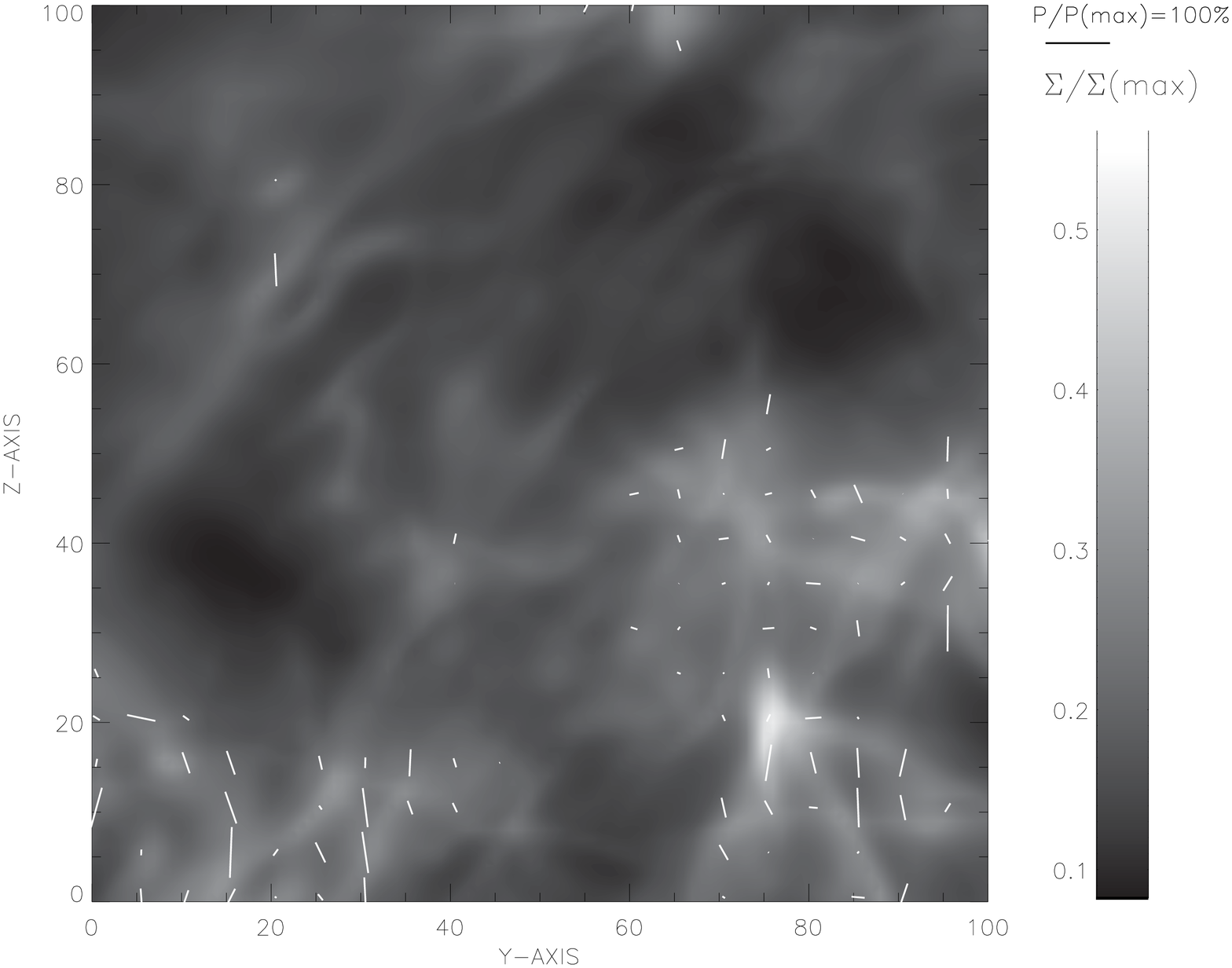}
   \hspace{0.2cm}
   \includegraphics[width=5cm]{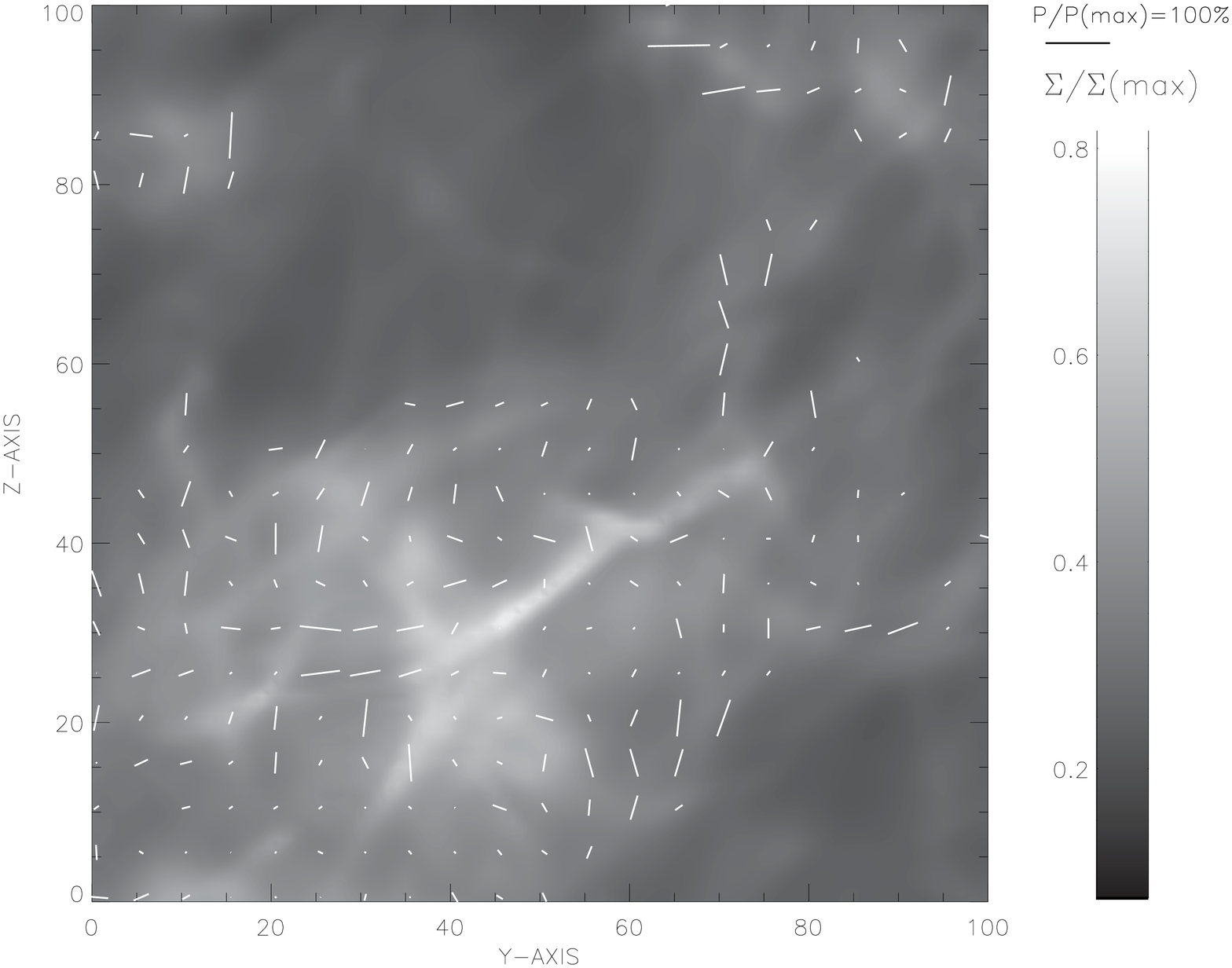}
   \caption{Polarization of emission and column density maps for Model 4 ($M_{\rm S} \sim 7.0$  and $M_{\rm A} \sim 7.0$) with ${\bf B}_{ext}$ perpendicular to the line of sight. The complete map (512x512 pixels) ({\it Upper-left}) and the zoomed regions (100x100 pixels). The sensitivity in simulated observations is assumed to be 0.3 of the maximum emission. Here, regions where the signal is less than 0.3 do not show polarization vectors, and $P_{\rm max} = 97\%$.}
\end{figure*}

\noindent
where $\rho$ is the local density and $i$ is the inclination of the local magnetic field and the line of sight.
We then obtain the integrated $Q$ and $U$, as well as the column density, along the LOS. 
Notice that, for the given equations the total intensity (Stokes $I$) is assumed to be simply proportional 
to the column density. The polarization degree is calculated from 
$P = \sqrt{Q^2+U^2}/I$ and the polarization angle $\phi = a\tan(U/Q)$. 

In Figs. 1 and 2 (see Falceta-Gon\c calves, Lazarian \& Kowal [2008], FLK08 hereafter), 
we show the obtained maps of column density and the polarization 
vectors for Models 3 and 4, respectively, for a line of sight perpendicular to the 
original magnetic field orientation. In Fig. 1, the polarization vectors are mostly 
uniform, and parallel to the external magnetic field. Here, fluctuations on the 
polarization angle are seen within the condensations, where the kinetic pressure 
overcomes the magnetic pressure. 
In Fig.\ 2, we show the column density and polarization maps of the super-Alfvenic case 
(Model 4). Here, the kinetic energy is larger than the magnetic pressure 
almost everywhere. As a consequence the gas easily tangles the magnetic 
field lines. The angular dispersion is larger and the polarization degree is smaller 
when compared to the sub-Alfvenic case. For the super-Alfvenic case, the orientation of the magnetic field 
regarding the LOS is irrelevant to the polarization maps. The strongly 
magnetized cases (Models 1, 2 and 3) present very similar polarization maps. Also, we 
must note that for LOS parallel to the original magnetic field, the polarization maps 
seem similar to the super-Alfvenic case.

Regarding the polarization degree, observations have shown that a decorrelation between 
the polarization degree and the column densities is detected for several objects 
\citep{matthews02, lai03, wolf03}. Typically, the polarization degree follows the 
relation $P \propto I^{-\gamma}$, where $\gamma \sim 0.5 - 1.2$ 
\citep{goncalves05} and $I$ is the total intensity. Usually, it is assumed that the grain 
alignment efficiency may be the general cause of this effect. 
However, as noticeable from Figs. 1 and 2, the polarization degree is smaller within high 
column density regions for all models. 
Note that we assumed perfect grain alignment, independent on the local density. 
Therefore, the decorrelation in our simulations must have a different cause. Another 
possibility could be the effect of the averaging along the LOS. Denser regions present 
a less uniform magnetic field. The annihilation of polarization vectors perpendicular to 
each other could cause the decrease in the polarization degree.

In Fig. 3 we show the correlation between the polarization degree and the column density 
for the different models, assuming that the magnetic field is in the plane of the sky. 
For all models, the polarization of 
high column densities tend to decrease to the minimum value ($\sim 20\% P_{\rm max}$). 
This minimum polarization degree should be zero for 
homogeneous density and random magnetic field. In inhomogeneous media it depends on the number of dense structures 
intercepted by the line of 
sight. The major contribution for the polarized emission comes from
dense clumps, which only are few along 
the LOS. This poor statistics 
results in a non-zero polarization degree. 
We also noticed that in subsonic turbulence the polarization 
degree is large even for the higher column densities. It occurs because in the subsonic 
models the contrast in density is small and the simulated domain is more homogeneous. 
Also, the number of dense clumps, which are able to tangle the field lines, is reduced in the sub-sonic case. 
On the contrary, for the super-Alfvenic case we obtain the minimum polarization degree 
everywhere, i.e.\ purely random magnetic field components.
For the decreasing part of the plots, we obtained a best fit with a correlation 
exponent $\gamma = 0.5$. If grain alignment was properly implemented in the calculations, 
the value of $\gamma$ should increase.
 
\begin{figure}
   \centering
   \includegraphics[width=\columnwidth]{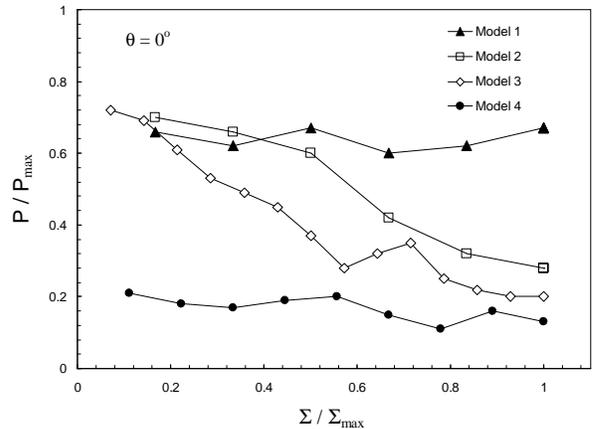} 
   \caption{Correlation between averaged 
polarization degree and the column density for the different models 
(FLK08). $P_{\rm max}$ is 100\%, 98\%, 97\% and 85\% for Model 1, 2, 
3 and 4, respectively (FLK08).}
\end{figure}

\subsection{Statistics of Polarization Angles}
\label{sec:math}

The histograms of polarization angles are shown in Fig.\ 4. In the upper panel we show the histograms for the 
sub-Alfvenic (Models 1, 2 and 3) and the super-Alfvenic (Model 4) cases, with the mean magnetic field lines perpendicular to the LOS. 
The polarization angles present very similar distributions and almost equal dispersion for the sub-Alfvenic cases. This happens mainly because they do not depend on the density structures, but on the 
magnetic topology. Strongly magnetized turbulence creates more filamentary and smoother density structures (i.e.\ low density contrast) if 
compared to weakly magnetized models and, most importantly the magnetic field lines are not highly perturbed. For Model 4, the distribution is practically homogeneous, which means that the polarization is randomly oriented in the plane of sky. It occurs because the turbulent/kinetic pressure is dominant and the gas is able to easily distort the magnetic field lines.

In the bottom panel of Fig.\ 4 we show the polarization angle histograms obtained for Model 3 
but for different orientations of the magnetic field. The dispersion of the polarization angle is very similar for 
inclination angles $\theta < 60^{\circ}$, and increases for larger inclinations. 
It may be understood if it is noted that the projected magnetic field $B_{sky} = B_{ext} \cos \theta$ is of order of the random component $\delta B$. 
It shows that the dominant parameter that differ the distributions of $\phi$ is the uniform magnetic field projected in the plane of sky, and not the intensity of the global magnetic field.

\begin{figure}
   \centering
   \includegraphics[width=\columnwidth]{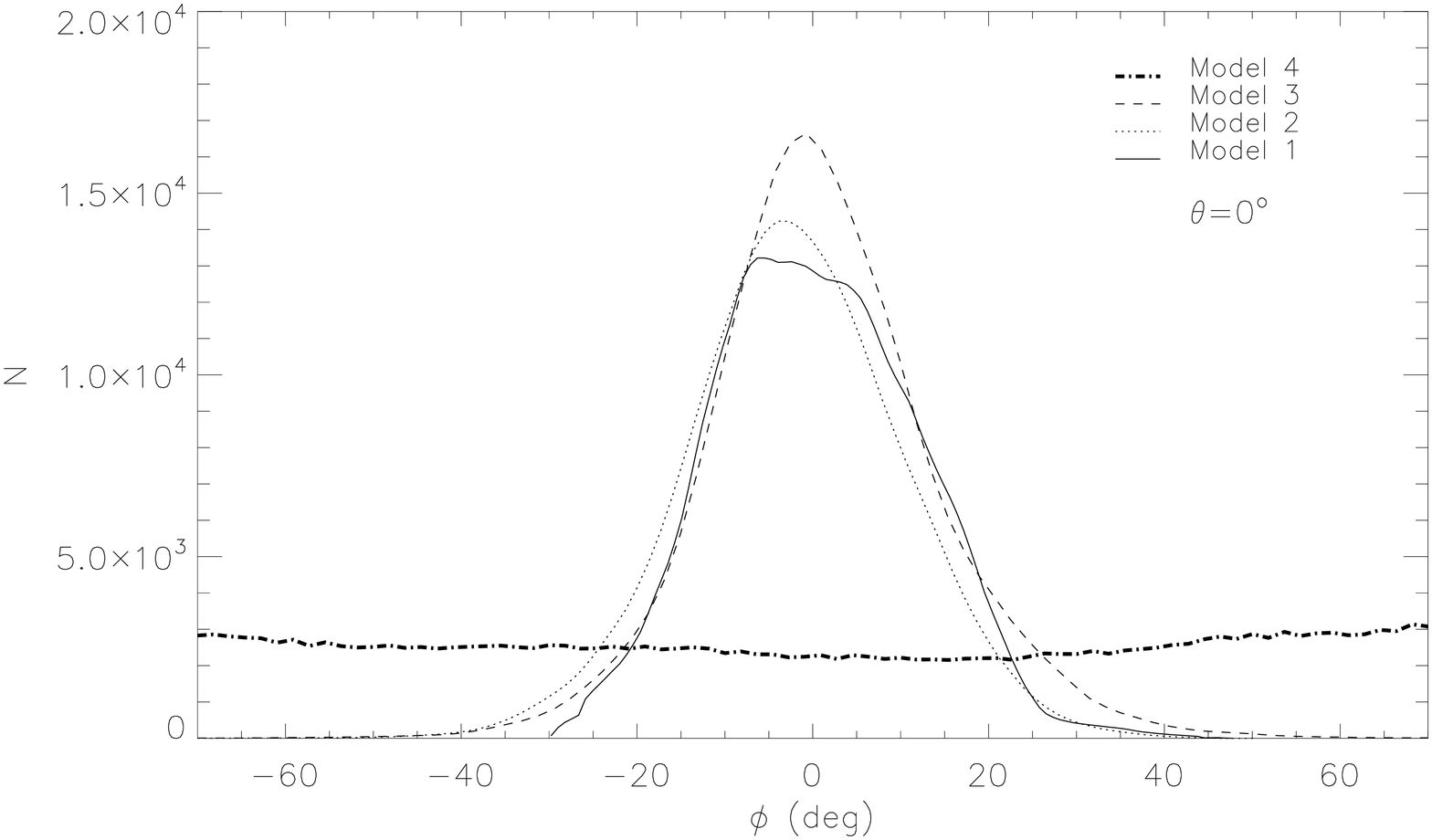}\\
   \includegraphics[width=\columnwidth]{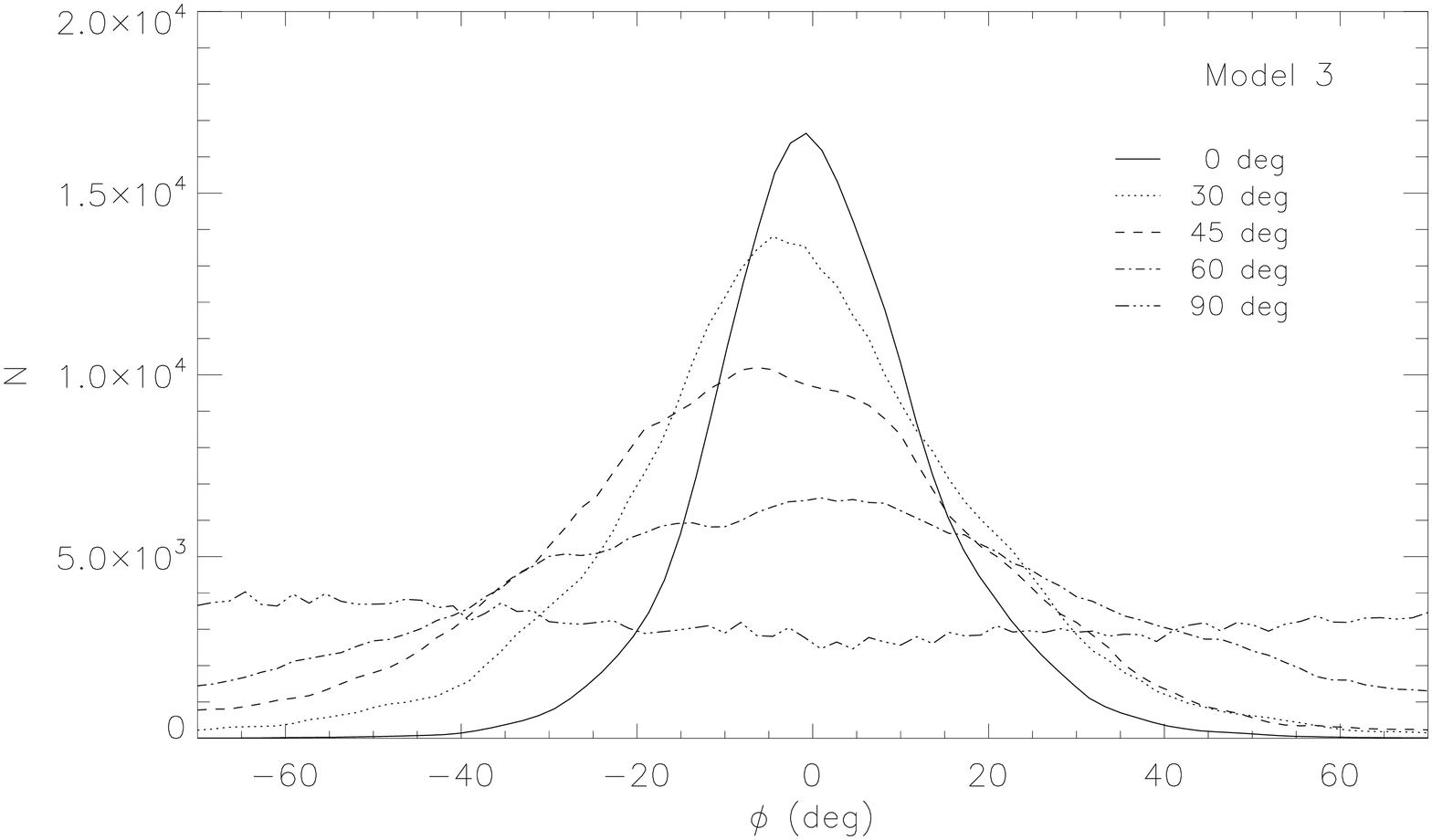}
   \caption{Histograms of polarization angle of the different models with $\theta=0$ ({\it up}), 
and for Model 3 and different magnetic field orientations in respect to the line of sight 
(angles $\theta$) ({\it bottom}) (FLK08).}\end{figure}

Another statistical analysis, which helps us to understand the topology of the 
polarization angles, is based on structure 
functions. The second order structure function (SF) of the polarization angles is 
defined as the average of the squared difference between the polarization angle measured at 2 points separated by a distance $l$:

\begin{equation}
{\rm SF}(l) = \left< \left| \phi \left({\bf r}+{\bf l }\right) - \phi \left( {\bf r} \right) \right|^2 \right>.
\end{equation}

The structure functions calculated for the different models are shown in Fig.\ 5. In the 
upper panel we present the SFs 
obtained for Model 3 with different values of $\theta$. As expected, all curves present a positive slope 
showing the increase in the difference 
of polarization angle for distant points. However, the small scales part of the SF presents 
a plateau extending up to $l \sim 4 - 5$ pix. This range corresponds to the dissipation 
region and may also be related to the smallest turbulent cells.
Surprisingly, the log-log slopes are equal independently on the
inclination of the LOS (for l between 3 and 20 grid points). The same 
behavior was obtained for all models. 
In the bottom panel we show the SFs calculated for 
the different 
models with $\theta = 0$. It is noticeable the increase in the SF for higher Mach numbers. However, the slopes are 
notably different. The maximum slope is $\alpha \sim 1.1$, 0.8, 0.5 and 0.3 for Models 1, 2, 3 and 4, respectively. 
Observationally, the molecular cloud M17 shows $\alpha \sim 0.5$ up to $l = 3$pc 
\citep{dotson96}, which would be in agreement with a cloud 
excited by supersonic and sub(or critically)-Alfvenic turbulence. 

\begin{figure}
   \centering
   \includegraphics[width=\columnwidth]{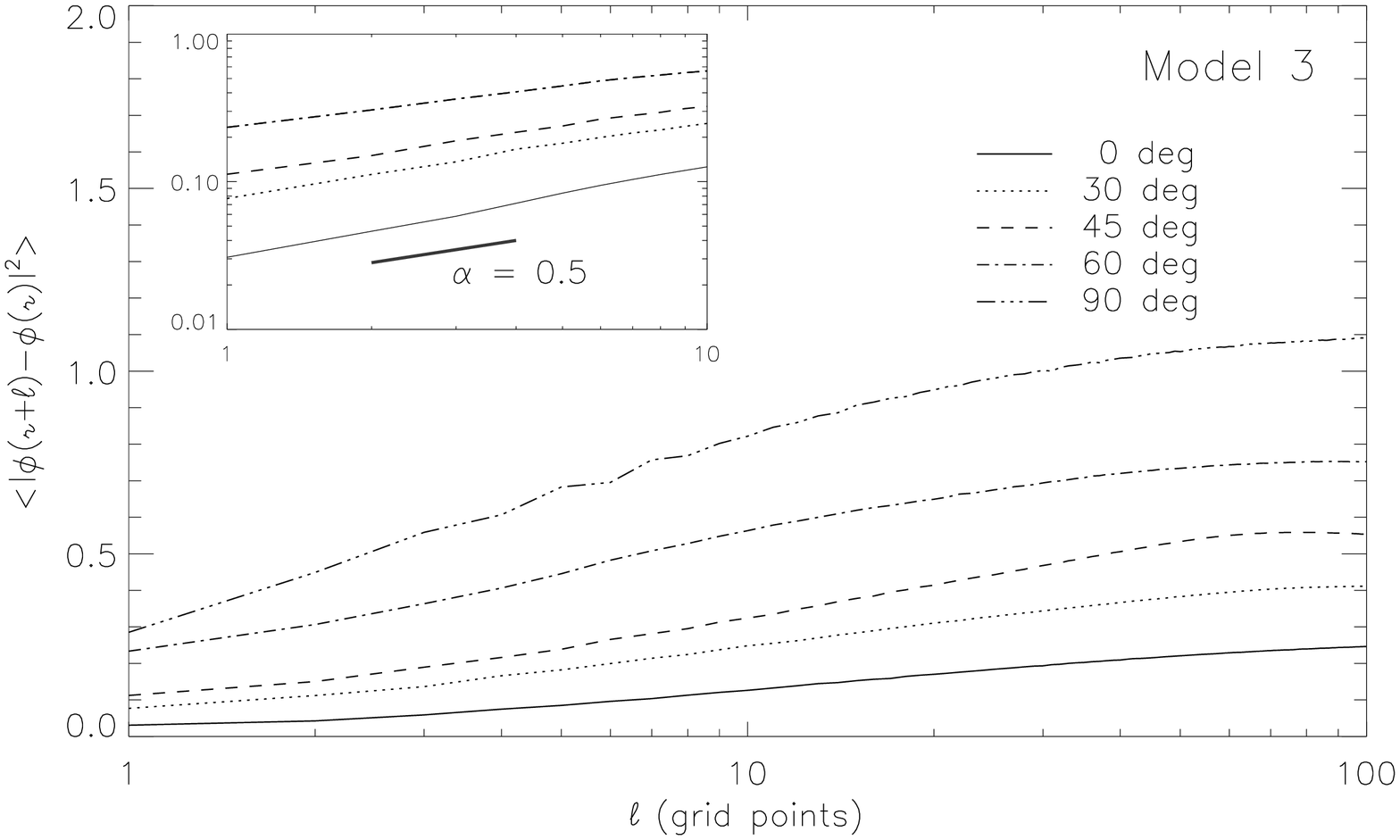}\\
   \includegraphics[width=\columnwidth]{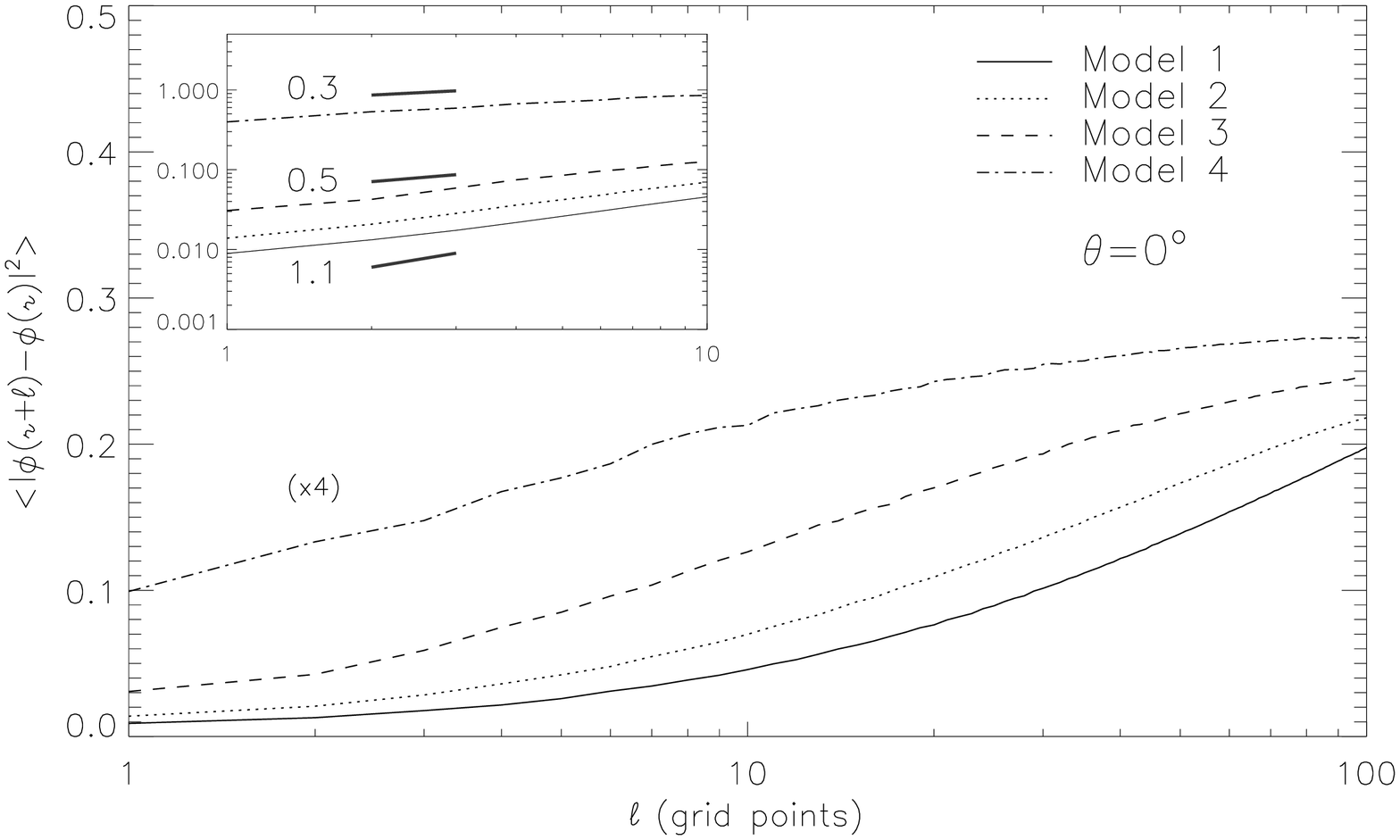}
   \caption{Structure functions of polarization angle for Model 3 with different magnetic field orientations 
regarding the line of sight (angles $\theta$) ({\it up}), and for the different 
models with magnetic field perpendicular to the line of sight ($\theta = 0$) ({\it 
bottom})(FLK08).}
\end{figure}

\subsection{The Chandrasekhar-Fermi Method}
\label{sec:math}

Chandrasekhar \& Fermi(1953) proposed a method to estimate the ISM magnetic fields based on the dispersion of polarization 
angles and the rms velocity. Basically, it is assumed that the magnetic field 
perturbations are Alfvenic, i.e.\ $\delta v \propto \delta B \sqrt\rho$. If the rms is 
isotropic, we may relate both as:

\begin{equation}
B_u \sim \frac{1}{2} \sqrt{4 \pi \rho} \ \frac{\delta V_{\rm LOS}}{\delta \phi},
\end{equation}

If one wants to expand the applicability of the CF method for cases where the random component of the magnetic 
field is comparable to the uniform component, or for larger inclination angles, it is necessary to take into account 
two corrections.
Firstly, we must introduce the total magnetic field projected in the plane of sky $B_{\rm sky} 
\sim B_{\rm sky}^{\rm ext} + \delta B$, where $B_{\rm sky}^{\rm ext}$ represents the mean field component projected 
on the plane of sky. We assume here, for the sake of simplicity, that $\delta B$ is 
isotropic. We assume that the $\delta B/B$ is a global relation and, in this case, 
we may firstly obtain the dispersion of $\phi$ and then calculate its tangent. 
Substituting $\delta \phi$ in Eq.\ (6) by $\tan(\delta \phi) \sim \delta B/B_{\rm sky}$, 
we obtain the modified CF equation: 

\begin{equation}
B_{\rm sky}^{\rm ext} + \delta B \simeq \sqrt{4 \pi \rho} \ \frac{\delta V_{\rm 
los}}{\tan \left(\delta \phi \right)}.
\end{equation} 

Since the CF-method strongly depends on the resolution (bacause of the dispersion of 
angles and velocities), we applied Eq.\ 
(7) to our simulated clouds, taking into account the effects of finite resolution. We 
calculated the 
average of the density weighted rms velocity along the LOS ($\delta V_{\rm los}$) and the dispersion of the polarization 
angle ($\delta \phi$) within regions of $R \times R$ pixels. To simulate a 
real cloud we substitute the parametric values of the Model 3 for $n_{\rm H} = 
10^3$cm$^{-3}$ and $T=10$K, and $B \sim 50\mu$G. The results showed a correlation between 
the obtained magnetic field from Eq. \ (7) and the resolution, which may be described by:

\begin{equation}
B_{\rm CF} = B_{\rm CF}^0 \left( 1+\frac{C}{R^{0.5}}\right),
\end{equation}

\noindent
where $R$ represents the observational resolution (total number of pixels), $C$ and $B_{\rm CF}^0$ are constants obtained from the best fit. $B_{\rm CF}^0$ represents the value of 
$B_{\rm CF}$ for infinite resolution observations, i.e.\ the best
magnetic field estimation from the CF method. 
%Eq.\ (10) is shown as the dotted lines in Fig.\ 9.

The fit parameters, as well as the 
expected values of the magnetic field from the simulations for all models, are shown in Table 2. Here, the 
magnetic fields are given in units of the mean field $B_{\rm ext}$. Since the simulations are scale independent, 
one could choose values of $B_{\rm ext}$ to represent a real cloud, in accordance with the parameters of Table 1. As an 
example, assuming a cloud with $n_{\rm H} = 10^3$cm$^{-3}$ and $T=10$K, and $\beta = 0.01$ (Model 3), we get 
$B \sim 50\mu$G. Choosing differently the density, temperature or the model given by the simulations, i.e.\ $\beta$,
 we obtain a different mean magnetic field. 
The obtained parameter $C$ is very similar for the different inclinations, but are different depending 
on the model, mainly because it is related to the scale on which the dispersion of the polarization angle changes. Since 
$C$ seems to depend on the model and not on the inclination it could also be used by observers to infer the physical 
properties of clouds from polarization maps.

\begin{table*}
\begin{center}
\caption{CF method estimates \label{table2}}
\begin{tabular}{cccccc}
\hline\hline
Model & $\theta (^{\circ})$ & $C$ & $B_{\rm CF}^0/B_{\rm ext}$ & $B_{\rm sky}^{\rm 
ext}/B_{\rm ext}$ & $B_{\rm tot}/B_{\rm ext}$ \\
3 & 0 &$20\pm5$ &$1.24 \pm 0.09$ & 1.00 & 1.25\\
3 & 30 &$24\pm5$ &$0.98 \pm 0.08$ & 0.87 & 1.11\\
3 & 45 &$25\pm5$ &$0.78 \pm 0.07$ & 0.71 & 0.96\\
3 & 60 &$33\pm5$ &$0.48 \pm 0.05$ & 0.50 & 0.75\\
3 & 90 &$31\pm5$ &$0.26 \pm 0.03$ & 0.00 & 0.24\\
\hline
1 & 0 &$7\pm5$ &$0.97 \pm 0.08$ & 1.00 & 1.11\\
2 & 0 &$10\pm5$ &$1.07 \pm 0.07$ & 1.00 & 1.16\\
4 & 0 &$34\pm5$ &$1.18 \pm 0.07$ & 1.00 & 1.41\\
\hline\hline
\end{tabular}
\end{center}
\end{table*}

As a practical use, observers could obtain polarimetric maps of a given region of the sky for different observational 
resolutions (e.g.\ changing the resolution via 
spatial averaging). Using the CF technique for each resolution and, then 
apply Eq.\ (8) to determine the asymptotic value of the magnetic field projected into the 
plane of sky $B_{\rm CF}^0$. 

\section{Conclusions}

In this work we presented turbulent 3-D high resolution MHD numerical simulations in order to study the 
polarized emission of dust particles in molecular clouds. We obtained synthetic dust 
emission polarization maps calculating the Stokes parameters $Q$, $U$ and $I$ 
assuming  a perfect grain alignment and that the dust optical properties are the same at all cells. 
Under these conditions, we were able to study the polarization angle distributions and the polarization degree for the 
different models and for different inclinations of the magnetic field regarding the LOS. 

As main results, we obtained an anti-correlation between the polarization degree and the 
column density, which is in agreement with 
observations, with exponent $\gamma \sim -0.5$. This value is related to random 
cancellation of polarization vectors integrated along the LOS, while larger indices 
require extra physics, e.g. dependency of the dust alignment efficiency with the local 
density. 

We showed that the overall properties of the polarization maps are related to the 
Alfvenic Mach number and not to the magnetic to gas pressure ratio. 
Also, we studied the PDF's and structure functions of the polarization angles, which 
showed a degeneracy of the results with the Alfvenic Mach number and the angle 
between the magnetic field and the LOS. Zeeman measurements of dense clouds may be useful 
to help remove this degeneracy as it could provide the magnetic field component along the 
LOS and, therefore, the inclination.

Finally, we presented a generalization of the CF method, which was
showed to be  useful for: (i) the 
determination of the total magnetic field projected in the plane of sky, and (ii) the 
separation of the two components $B_{\rm sky}$ and $\delta B$;

\acknowledgments

D.F.G., A.L. and G.K. thank the financial support of the NSF (No.\ AST0307869), the Center for Magnetic Self-Organization 
in Astrophysical and Laboratory Plasmas and the Brazilian agencies FAPESP (No.\ 06/57824-1 and 07/50065-0) and CAPES (No.\ 4141067).

\end{document}